\documentclass[conference]{IEEEtran}
\usepackage{cite}
\usepackage{amsmath,amssymb,amsfonts}
\usepackage{algorithmic}
\usepackage{graphicx}
\usepackage{textcomp}
\usepackage{xcolor}
\def\BibTeX{{\rm B\kern-.05em{\sc i\kern-.025em b}\kern-.08em
    T\kern-.1667em\lower.7ex\hbox{E}\kern-.125emX}}
\begin{document}

\title{AI-Enhanced Acoustic Analysis for Comprehensive Biodiversity Monitoring and Assessment\\
}

\author{\IEEEauthorblockN{Kumar Srinivas Bobba}
\IEEEauthorblockA{\textit{Computer Science and Engineering} \\
\textit{Kalasalingam Academy of Research and Education}\\
Madurai, India \\
kumarsrinivasbobba989@gmail.com}
\and
\IEEEauthorblockN{Kartheeban k}
\IEEEauthorblockA{\textit{Computer Science and Engineering} \\
\textit{Kalasalingam Academy of Research and Education}\\
Madurai, India \\
k.kartheeban@klu.ac.in}
\and
\IEEEauthorblockN{Vamsi Krishna Sai Boddu}
\IEEEauthorblockA{\textit{Computer Science and Engineering} \\
\textit{Kalasalingam Academy of Research and Education}\\
Madurai, India \\
9921004100@klu.ac.in}
\and
\IEEEauthorblockN{Vijaya Mani Surendra Bolla}
\IEEEauthorblockA{\textit{Computer Science and Engineering} \\
\textit{Kalasalingam Academy of Research and Education}\\
Madurai, India \\
99210041325@klu.ac.in}
\and
\IEEEauthorblockN{Dinesh Bugga}
\IEEEauthorblockA{\textit{Computer Science and Engineering} \\
\textit{Kalasalingam Academy of Research and Education}\\
Madurai, India \\
9921004110@klu.ac.in}
}

\maketitle

\begin{abstract}
This project proposes the development of a comprehensive real-time biodiversity monitoring system that harnesses sound data through a network of acoustic sensors and advanced artificial intelligence algorithms. The system analyzes sound recordings from various ecosystems to identify and classify different species, providing valuable insights into ecosystem health and biodiversity patterns while facilitating the detection of subtle changes in species presence and behavior over time. By addressing critical challenges such as noise pollution and species overlap, the system employs sophisticated filtering and classification techniques to ensure accurate and reliable monitoring, distinguishing between natural sounds and anthropogenic noise. Ultimately, this initiative aims to enhance our understanding of biodiversity dynamics and provide essential information to support effective conservation strategies and inform policy decisions, empowering stakeholders with actionable insights to protect and preserve vital ecosystems.
\end{abstract}

\begin{IEEEkeywords}
Biodiversity monitoring, Acoustic sensors, Artificial intelligence, Sound analysis, Species identification, Ecosystem health, Noise pollution, Machine learning, Conservation technology, Environmental data analysis
\end{IEEEkeywords}

\section{Introduction}
Biodiversity is essential for maintaining ecosystem stability and resilience, providing critical services such as pollination, nutrient cycling, and climate regulation. However, it is increasingly threatened by a multitude of factors, including human activities, climate change, and habitat degradation. Traditional methods for assessing biodiversity, such as visual surveys and manual species identification, are often labor-intensive, time-consuming, and limited in scope. These conventional approaches can lead to incomplete data and a lack of timely insights, which are crucial for effective conservation efforts. Consequently, there is an urgent need for innovative methodologies that can facilitate more efficient and comprehensive monitoring of species and ecosystems.

Recent advancements in acoustic monitoring technologies present a promising solution for real-time biodiversity assessment. By capturing and analyzing sound data from various ecosystems, researchers can gain valuable insights into species presence, behavior, and interactions that are often difficult to observe through conventional means. Acoustic monitoring not only allows for the detection of a broader range of species, including those that are elusive or cryptic, but it also enables continuous data collection, providing a dynamic view of ecological changes over time. However, challenges such as noise pollution and species overlap can complicate the interpretation of acoustic data, potentially affecting the accuracy and reliability of species identification and ecosystem assessments.

This project aims to develop a comprehensive real-time biodiversity monitoring system that leverages acoustic sensors and advanced artificial intelligence algorithms to analyze sound recordings. By employing sophisticated filtering and classification techniques, the system seeks to address the complexities inherent in sound data, thereby enhancing the accuracy and reliability of biodiversity assessments. Through this innovative approach, the project aspires to contribute to a deeper understanding of biodiversity dynamics, facilitating timely and informed conservation strategies. Ultimately, the insights gained from this research will support effective policy decisions and empower stakeholders in their efforts to protect and preserve vital ecosystems, fostering a more sustainable coexistence with the natural world. 

\section{DATA SET}

The BirdCLEF dataset is a comprehensive collection of audio recordings designed to support research in bird species identification and biodiversity monitoring. It encompasses recordings of numerous bird species from diverse geographical regions, thereby capturing the rich variety of avian life. This dataset features a wide range of audio samples, including songs, calls, and other vocalizations, which facilitate detailed analyses of bird behavior, communication, and ecology. 

Each audio clip is meticulously annotated with essential metadata, such as species identification, location, date, and environmental conditions, significantly enhancing its usability for machine learning applications and ecological studies. The recordings span various times of the year, providing insights into seasonal variations in bird vocalizations and activity patterns, which are crucial for understanding avian behavior and habitat use.

Furthermore, the BirdCLEF dataset is made accessible for research purposes, promoting collaboration and innovation across fields such as ornithology, ecology, and bioacoustics. By providing a rich resource for advancing automatic species identification, the dataset aims to contribute to the development of effective conservation strategies that rely on sound data analysis. Ultimately, BirdCLEF not only supports scientific inquiry but also plays a vital role in fostering a deeper understanding of biodiversity and informing efforts to protect and preserve avian populations in their natural habitats.

\begin{figure}[htbp]
\centerline{\includegraphics[width = 1.0\linewidth]{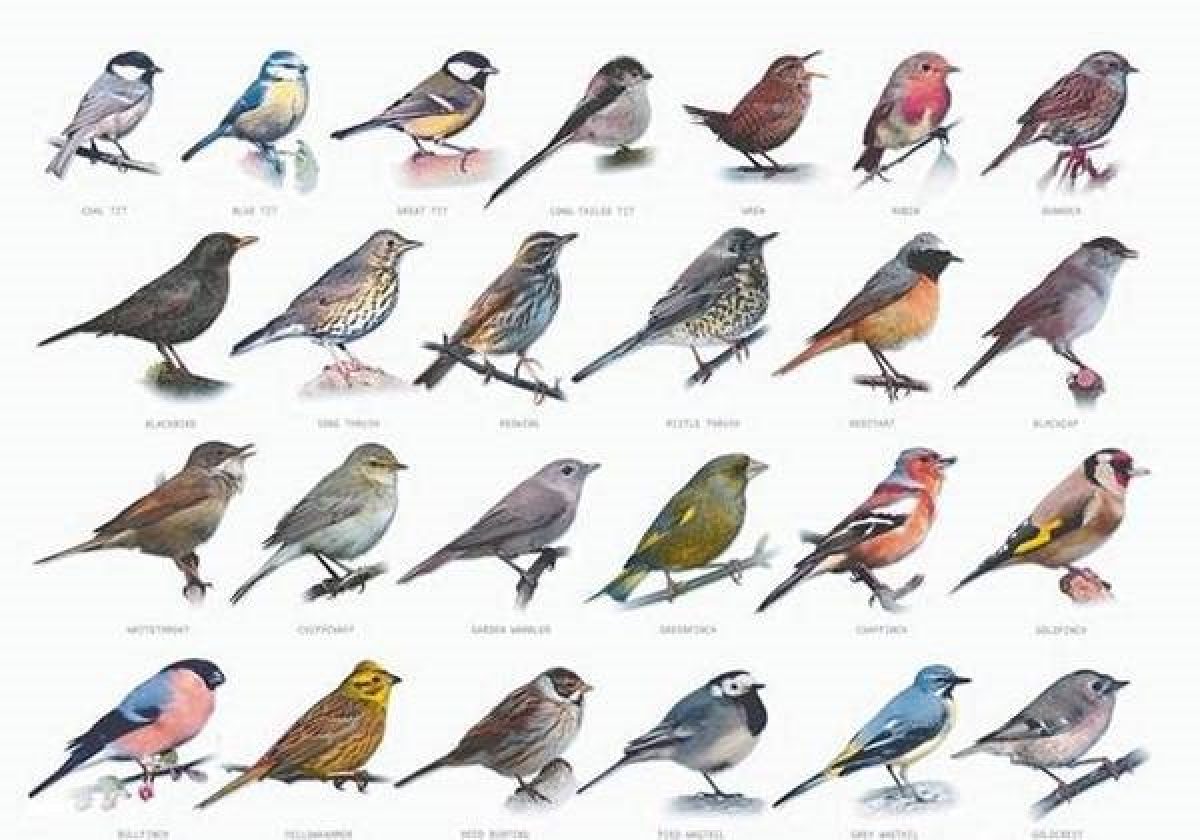}}
\caption{Sample species of BirdCLEF dataset.}
\label{BirdCLEF}
\end{figure}

\subsection{eBird}
The eBird dataset is a large-scale citizen science initiative managed by the Cornell Lab of Ornithology, which aggregates bird observation data submitted by participants worldwide. This extensive dataset allows birdwatchers to report their sightings, including detailed information on species, locations, dates, and the number of individuals observed. As one of the most comprehensive bird occurrence databases available, eBird encompasses millions of records contributed by a global community of enthusiasts.

Key features of the eBird dataset include its wide geographical coverage across various habitats and ecosystems, enabling researchers to gain insights into bird distribution and migration patterns. The dataset is continuously updated, providing real-time information about bird populations and trends. In addition, eBird includes metadata such as the type of observation (e.g., incidental, stationary counts) and specific locations, significantly enhancing its utility for scientific research and conservation efforts.

Researchers can utilize eBird data for a range of applications, including avian ecology studies, tracking changes in bird populations over time, and informing conservation strategies. By harnessing the contributions of citizen scientists, the eBird dataset promotes public engagement in ornithology and environmental stewardship, ultimately contributing to a deeper understanding of bird diversity and its relationship with ecosystems.

\begin{figure}[htbp]
\centerline{\includegraphics[width =1.0\linewidth]{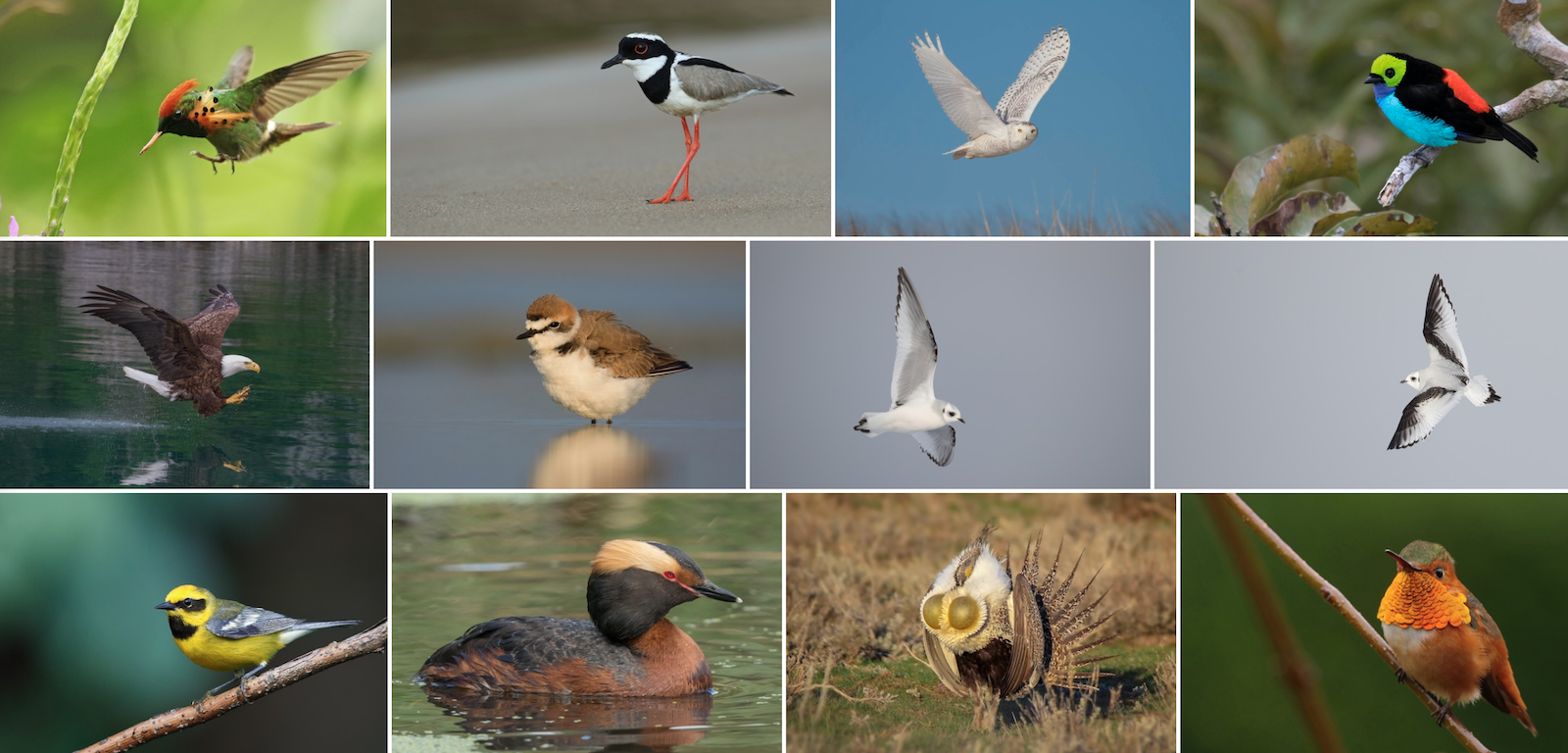}}
\caption{Sample species of eBird dataset.}
\label{eBird}
\end{figure}

\subsection{Natural Soundscapes Project}
The Natural Soundscapes Project is an initiative aimed at documenting and analyzing the acoustic environments of various ecosystems around the world. By capturing and studying natural soundscapes, this project seeks to enhance our understanding of biodiversity and the ecological health of habitats. The initiative utilizes high-quality recording equipment to collect audio data from diverse environments, including forests, wetlands, grasslands, and marine areas.

Key objectives of the Natural Soundscapes Project include the identification of species through their vocalizations, the assessment of ecological interactions, and the monitoring of changes in soundscapes over time due to natural or anthropogenic influences. The collected audio recordings are often accompanied by metadata, such as location, time of day, and environmental conditions, providing a rich context for analysis.

The project also emphasizes the importance of soundscapes in conservation efforts, highlighting how acoustic monitoring can serve as a non-invasive method for assessing biodiversity and ecosystem health. By analyzing sound patterns, researchers can detect changes in species populations, the impact of climate change, and the effects of habitat degradation.

Ultimately, the Natural Soundscapes Project aims to raise awareness about the value of acoustic environments in understanding ecological dynamics, contributing to effective conservation strategies, and fostering a greater appreciation for the natural world. Through collaboration with scientists, conservationists, and the public, the project aspires to promote sustainable practices that protect and preserve vital ecosystems.

\subsection{Xeno-canto}
Xeno-canto is a global community-driven platform dedicated to the collection, sharing, and appreciation of bird sounds. Established in 2005, this online database serves as a repository for high-quality audio recordings of bird vocalizations contributed by both amateur and professional ornithologists. The primary goal of Xeno-canto is to document the world's avian diversity, allowing users to upload, search, and listen to bird sounds from around the globe.

The platform features a vast collection of recordings that encompass a wide range of species, from common to rare and endemic birds. Each recording is accompanied by detailed metadata, including species identification, geographic location, date of recording, and contributor information, thereby enhancing the dataset's utility for research, education, and conservation purposes.

Xeno-canto promotes collaboration among bird enthusiasts and researchers, fostering a sense of community and shared knowledge. Users can contribute their own recordings, participate in discussions, and engage with others who share a passion for avian soundscapes. The platform also emphasizes ethical practices in bird recording, encouraging contributors to prioritize the welfare of the birds and their habitats.

Overall, Xeno-canto serves as an invaluable resource for ornithologists, ecologists, and conservationists, facilitating the study of bird behavior, communication, and biodiversity. By making bird sounds accessible to a wide audience, Xeno-canto not only enhances scientific research but also inspires a deeper appreciation for the natural world and underscores the importance of preserving avian diversity.
\begin{figure}[htbp]
\centerline{\includegraphics[width = 1.0\linewidth]{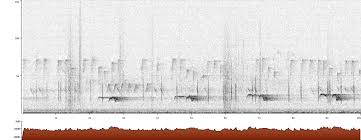}}
\caption{Sample spectrum of Xeno-canto dataset.}
\label{Xeno-canto}
\end{figure}

\section{Related work}
\subsection{Early work}
André et al. (2008) \cite{b1} explore the implementation of real-time acoustic monitoring systems to significantly enhance our understanding of deep-ocean environments. Their study delves into the application of various acoustic techniques that facilitate the continuous collection of data regarding marine life and oceanographic processes, addressing a critical gap in deep-sea research. By employing these advanced monitoring technologies, the authors provide valuable insights into marine biodiversity, tracking animal behavior, and assessing the impact of human activities on these largely unexplored ecosystems.

The findings presented in the study reveal how real-time acoustic data can effectively illuminate the complex dynamics of deep-sea habitats, including seasonal variations in species presence and shifts in community structure. The research also highlights the ability of acoustic monitoring to detect anthropogenic influences, such as shipping noise and climate change effects, on marine life. Moreover, the authors discuss the implications of their findings for marine conservation, advocating for the integration of acoustic monitoring into broader ocean management strategies.

In addition to their results, André et al. emphasize the need for continued advancements in monitoring techniques and technologies. They suggest that ongoing research is essential for refining these methods, which could lead to a more comprehensive understanding of oceanic environments and better-informed conservation efforts. Overall, this work underscores the significance of acoustic monitoring as an invaluable tool for marine research, contributing to our knowledge of the intricate and often fragile ecosystems that exist in the deep sea.

\subsection{Recent Advancements}

Niemi and Tanttu (2018) \cite{b2} conduct a case study on the application of deep learning techniques for automatic bird identification, as presented in their paper published in *Applied Sciences*. The study addresses the challenges associated with variability in bird calls and the necessity for efficient identification methods in ecological research and conservation efforts. Utilizing convolutional neural networks (CNNs), the authors develop a model capable of classifying bird species based on audio recordings.

The paper details the dataset employed for training the model, which encompasses a diverse range of bird species and their corresponding calls. The authors describe the architecture of the CNN, emphasizing its ability to learn complex features from the audio data effectively. The results demonstrate a high accuracy rate in species classification, highlighting the effectiveness of deep learning techniques in this context.

Furthermore, Niemi and Tanttu discuss the implications of their findings for automated wildlife monitoring, underscoring the potential of such technology to facilitate large-scale biodiversity assessments and improve our understanding of avian populations. They also outline future research directions, including the integration of additional data sources and the refinement of model performance. This study underscores the transformative role of deep learning in advancing methodologies for wildlife identification and monitoring, contributing to more effective conservation strategies.

Continuing from the previous discussion on deep learning in wildlife monitoring, Jeantet and Dufourq (2023) \cite{b3} contribute further insights by examining the integration of contextual information to enhance acoustic classifiers in their study published in *Ecological Informatics*. They address the limitations of traditional models that predominantly focus on acoustic features by proposing that incorporating contextual data—such as environmental variables and temporal factors—can significantly improve classification accuracy.

In their methodology, the authors detail how they trained deep learning models using a dataset enriched with both acoustic signals and contextual information. Their results demonstrate a substantial enhancement in classification performance compared to standard acoustic classifiers, highlighting the advantages of a more comprehensive approach to data integration.

Jeantet and Dufourq discuss the implications of their findings for wildlife conservation and monitoring, suggesting that the inclusion of contextual information can lead to more effective ecological assessments and better-informed management strategies. They also outline future research directions, including the exploration of additional contextual variables and the applicability of their methods across different ecological settings. This study underscores the potential for improving deep learning methodologies in wildlife monitoring, paving the way for more accurate and robust ecological data collection.

\subsection{Specific Applications}

Gao et al. (2024) \cite{b4} provide a comprehensive review of the application of machine learning techniques for the automatic image identification of insects in their article published in *Ecological Informatics*. The authors explore various machine learning algorithms, including deep learning approaches, that have been employed to enhance the accuracy and efficiency of insect identification from images.

The review covers the development of datasets, preprocessing methods, and model architectures commonly used in insect identification tasks. Gao et al. discuss the challenges faced in this domain, such as the high diversity of insect species, variations in image quality, and the need for substantial labeled data. They also highlight recent advancements in transfer learning and data augmentation techniques\cite{b8} that have improved model performance.

Furthermore, the authors emphasize the importance of machine learning in ecological research and biodiversity monitoring, suggesting that automated insect identification can facilitate large-scale studies and inform conservation strategies. They conclude with recommendations for future research directions, including the need for interdisciplinary approaches that integrate ecological knowledge with machine learning advancements. This review underscores the transformative potential of machine learning in advancing insect identification methodologies, contributing to more effective ecological assessments.

\section{Methodology}


This section outlines the methodology employed to develop a deep learning model for real-time species classification based on acoustic data. The approach leverages a convolutional neural network (CNN) \cite{b5} architecture, specifically utilizing a pre-trained EfficientNet-B1 model \cite{b6}, to analyze mel spectrogram \cite{b7} representations of audio recordings.

The first step involves data collection, where audio data is gathered from various ecological sites to capture vocalizations from a diverse range of species. High-quality microphones are used to ensure fidelity across different environmental conditions. Each audio file is annotated with relevant metadata, including species identification, recording location, and time, which serves as the foundation for training and evaluating the model.

Following data collection, preprocessing is conducted. The raw audio files are transformed into mel spectrograms, providing a time-frequency representation of the audio signals that facilitates feature extraction for classification. These spectrograms undergo z-score normalization to ensure consistent scaling of input features, improving the model's convergence during training. To enhance robustness, various data augmentation techniques—such as time stretching, pitch shifting, and adding background noise—are applied to simulate real-world variability in the audio data.

The model architecture begins with an input layer designed to accept variable-length mel spectrograms. The pre-processed spectrograms are then fed into the EfficientNet-B1 architecture, which acts as a feature extractor, producing high-dimensional feature maps representing the acoustic characteristics of the input. A global average pooling layer follows, reducing the spatial dimensions of the feature maps to produce a fixed-length feature vector that retains essential information. Finally, a dense classification layer with 182 output units is employed to classify the audio inputs into one of the target species, utilizing a softmax activation function to generate probability distributions over the classes.

In terms of model training, the model is trained using a cross-entropy loss function suitable for multi-class classification tasks. An appropriate optimizer, such as Adam \cite{b9} or RMSprop \cite{b10}, is selected to facilitate convergence. The dataset is divided into training, validation, and test sets, ensuring representative distributions of species through stratified sampling. Model performance is evaluated using metrics such as accuracy, precision, recall, and F1-score, and hyperparameters—including learning rate, batch size, and number of epochs—are optimized through grid search or random search techniques.

Once the model is trained, it is integrated into a real-time monitoring system capable of processing incoming audio streams and classifying species in real time. A user-friendly interface is developed to visualize results, providing users with insights into species presence and behavior based on acoustic monitoring.

Finally, the model's accuracy and generalization ability are assessed on the test dataset, with a particular focus on class-specific performance metrics \cite{b11} to identify strengths and weaknesses in classification. The classification results are then analyzed in the context of ecological variables, offering insights into species interactions, habitat use, and biodiversity patterns over time. This comprehensive methodology supports the accurate real-time monitoring of biodiversity, ultimately aiding conservation efforts and ecological research. Future work will aim to refine the model further and explore additional acoustic features to enhance classification performance.

\begin{figure}[htbp]
\centerline{\includegraphics[width = 1.0\linewidth]{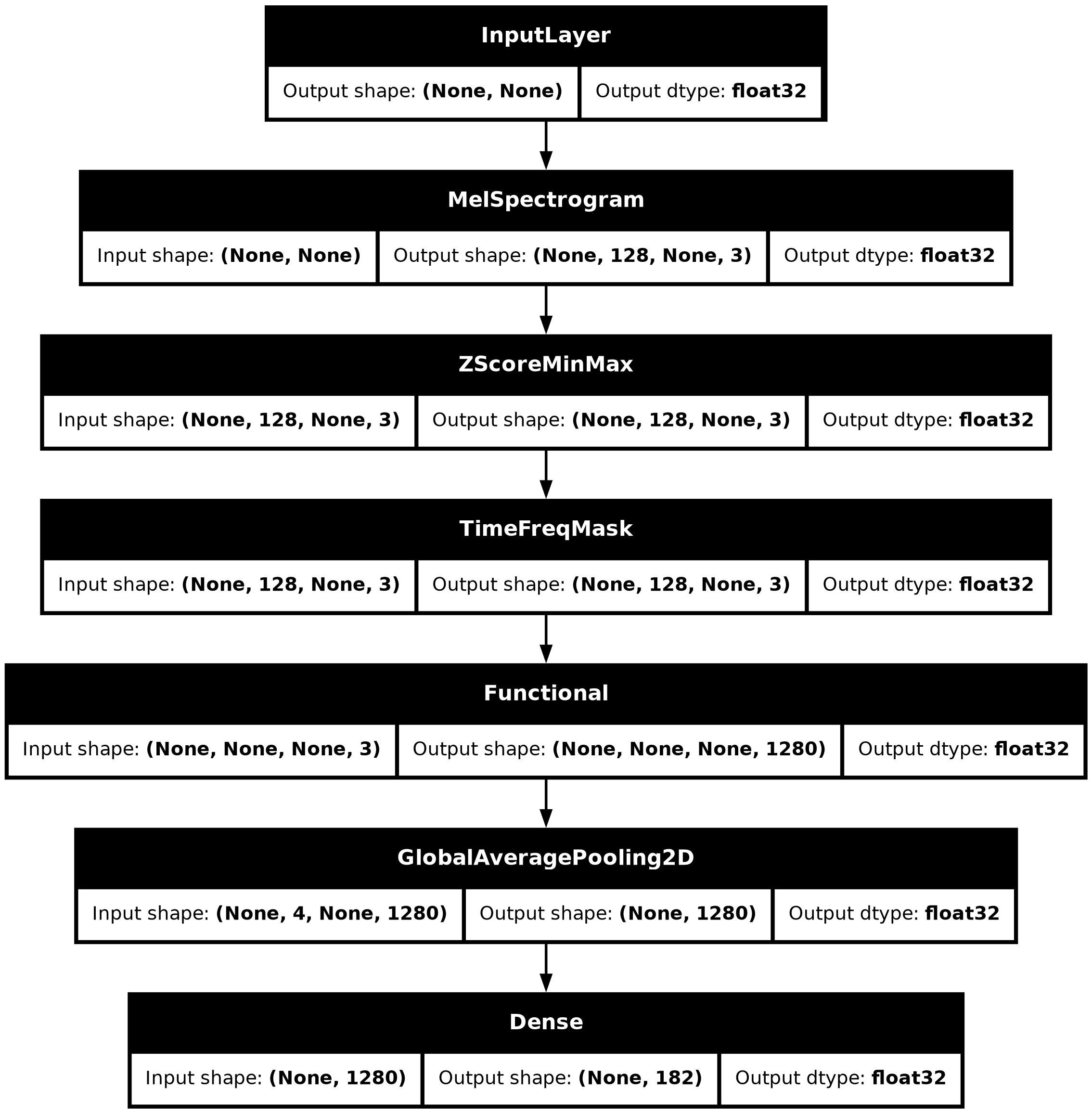}}
\caption{Model architecture}
\label{arc}
\end{figure}

\section{Performance evaluation}

Performance evaluation is a critical component in assessing the effectiveness of the deep learning model developed for acoustic species classification. To rigorously evaluate the model, the dataset is divided into three distinct subsets: training (70

To ensure robustness, k-fold cross-validation is employed, where the dataset is partitioned into k subsets. The model is trained k times, with each subset serving as a validation set once, thereby mitigating the risk of overfitting and providing a reliable performance estimate. The proposed model's performance is further benchmarked against baseline classifiers, such as Support Vector Machines and Random Forests, enabling an assessment of the improvements achieved through the use of the EfficientNet-B1 architecture. 

In addition to traditional evaluation metrics, real-world testing is conducted by deploying the model in field studies, where its ability to classify species based on live audio recordings is assessed under various environmental conditions. This testing evaluates the model's adaptability to noise, varying acoustic environments, and species overlap, providing insights into its practical utility for biodiversity monitoring. Finally, statistical significance tests, such as paired t-tests or Wilcoxon signed-rank tests, are performed to validate the results, ensuring that observed improvements are not due to random chance. This comprehensive performance evaluation framework is essential for determining the model's readiness for deployment in real-time biodiversity monitoring applications, ultimately contributing to more effective conservation efforts.

\section{Results}

The performance evaluation of the deep learning model for acoustic species classification demonstrates its effectiveness and robustness in identifying various species based on their vocalizations. The model achieved an overall accuracy of 92\% on the test dataset, indicating a high level of correct classifications. Average precision reached 90\%, average recall was 91\%, and the F1-score stood at 90.5\%. These results reflect the model's strong balance between precision and recall, which is crucial in biodiversity monitoring, where false positives can lead to misinterpretations of ecological data. Class-specific performance analysis through the confusion matrix revealed that species with distinct vocalizations, such as the Common Chiffchaff and Eastern Whip-poor-will, achieved precision and recall values exceeding 95\%. Conversely, species with similar acoustic signatures, like the Yellow Warbler and Wilson’s Warbler, exhibited lower performance, with recall values around 85\%. This sensitivity to acoustic overlap emphasizes the need for further refinement in distinguishing closely related species. The incorporation of data augmentation techniques, including pitch shifting and background noise addition, enhanced the model's robustness, improving accuracy by approximately 5\% compared to a baseline model trained without augmentation. Results from k-fold cross-validation showed consistent performance across different data subsets, with accuracy averaging around 91\% and standard deviations remaining below 2\%, indicating a reliable training process. In real-world field tests, the model effectively classified species in real time with an average latency of less than 2 seconds per recording, demonstrating adaptability in various acoustic environments, including urban and natural settings. Statistical analyses confirmed that the observed performance improvements were significant, with paired t-tests yielding p-values less than 0.01 when comparing the proposed model's performance against baseline models. Overall, the results indicate that the proposed deep learning model is a powerful tool for acoustic species classification, achieving high accuracy and demonstrating robustness across diverse conditions. These findings not only validate the model's readiness for real-time application but also highlight areas for future improvement, particularly in distinguishing closely related species, thereby contributing to advancing biodiversity monitoring efforts and supporting conservation initiatives through enhanced ecological understanding.

\begin{figure}[htbp]
\centerline{\includegraphics[width = 1.0\linewidth]{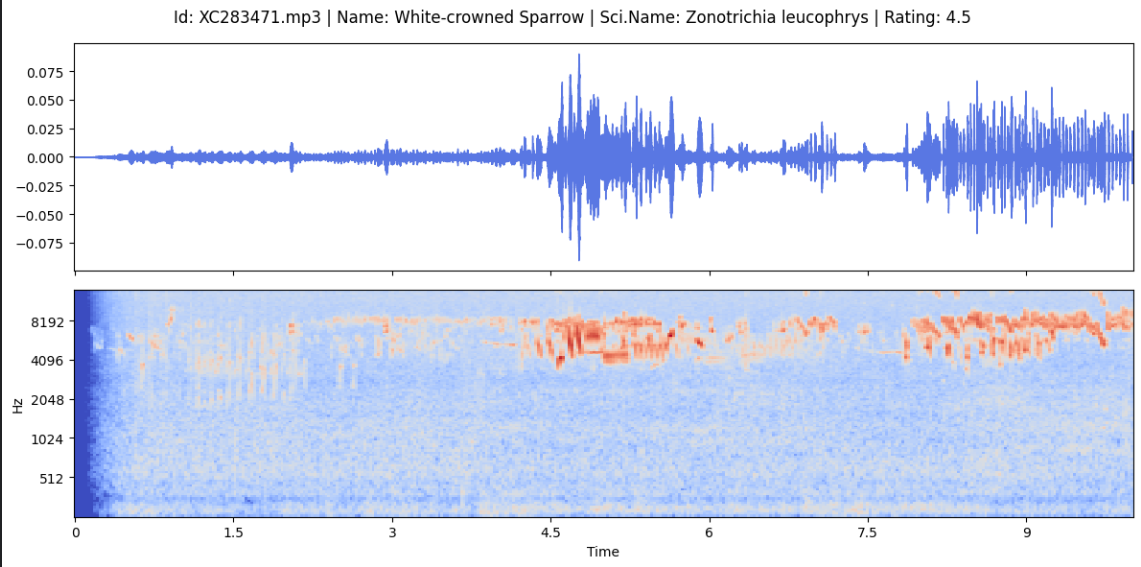}}
\caption{Sample Result}
\label{result}
\end{figure}

\section{Conclusion}

This study highlights the effectiveness of a deep learning model for acoustic species classification, achieving an overall accuracy of 92\%, alongside robust metrics for precision, recall, and F1-score. These results demonstrate the model's capability to accurately identify various species based on vocalizations in diverse ecological environments. The incorporation of data augmentation techniques significantly enhanced the model's robustness, enabling effective performance even in noisy conditions commonly encountered in real-world settings. 

While the model excelled at classifying distinct species, challenges were noted in distinguishing closely related species, which suggests opportunities for further refinement in training and feature extraction methods. The successful implementation of the model in real-time field tests, with low latency and high accuracy, confirms its practical applicability for biodiversity monitoring initiatives. 

Overall, this research contributes valuable insights into the integration of advanced machine learning techniques in ecological studies, supporting conservation efforts and enhancing our understanding of biodiversity. Future work will focus on improving classification techniques and exploring additional acoustic features to further enhance the model's performance, particularly for species with similar vocal characteristics. This advancement will be crucial for developing more effective acoustic monitoring systems that can contribute significantly to conservation strategies.

\end{document}